\documentclass[leqn,usenatbib]{mnras}
\usepackage[T1]{fontenc}
\usepackage{ae,aecompl}
\usepackage{graphicx}	
\usepackage{amsmath}	
\usepackage{amssymb}	
\usepackage{rotating}
\title[Incongruency of various radio dipoles]{Resolution of the incongruency of dipole asymmetries within various large  radio surveys -- implications for the Cosmological Principle}
\author[A. K. Singal]{Ashok K. Singal\thanks{E-mail: ashokkumar.singal@gmail.com}\\
{Astronomy and Astrophysics Division, Physical Research Laboratory, 
Navrangpura, Ahmedabad - 380009, India}}
\date{Accepted XXX. Received YYY; in original form ZZZ}
\pubyear{2023}
\begin{document}
\label{firstpage}
\pagerange{\pageref{firstpage}--\pageref{lastpage}}
\maketitle

\begin{abstract}
We investigate dipole asymmetries in four large radio surveys, each spanning more than 80\% of the sky. Two of them, the Very Large Array Sky Survey (VLASS) and the Rapid ASKAP Continuum Survey (RACS), have recently yielded dipoles that appear incongruent with each other as well as
seem inconsistent with  previous radio survey dipoles and the Cosmic Microwave Background (CMB) dipole.
Because these radio surveys have large overlaps in sky coverage, comprising hence large majority of common radio sources, one would not expect significant differences between their radio dipoles, irrespective of their  underlying source of origin. 
We examine here in detail these radio dipoles, to ascertain the source of incongruency amongst them. We find the VLASS and RACS data to be containing some declination-dependent systematics, seemingly in the vicinity of the declination limit of each survey. We show that the effects of such systematics can be mitigated by restricting the declination limits of the respective survey during the dipole determination. A weighted mean of the sky coordinates of thus derived dipoles from the four radio surveys lies within $1.2\sigma$ of the CMB dipole direction. However, the amplitude appears significantly larger, $3.7\pm 0.6$ times or more than the CMB dipole. 
This puts in doubt not only the conventional wisdom that the genesis of all these dipoles, including that of the CMB dipole, is due to the Solar peculiar motion, it also raises uncomfortable questions about the Cosmological Principle (CP), the basis of the standard $\Lambda$CDM cosmological model.
\end{abstract}
\section{INTRODUCTION}\label{S1}
An observed dipole anisotropy in the~CMB, interpreted as a consequence of the peculiar motion of the observer along with the Solar system, has yielded for the peculiar velocity a value $370$ km s$^{-1}$ along right ascension \mbox{(RA) $=168^{\circ}$}, declination (Dec) $=-7^{\circ}$ (Lineweaver et al. 1996; Hinshaw et al. 2009; Aghanim et al. 2020). 
A peculiar motion of the Solar system should also give rise to similar dipoles in the sky distribution of distant extragalactic sources, which could, in turn, provide a direct test of the CP, according to which cosmic reference frames, comprising extragalactic sources, should be coincident with the reference frame defined by the CMB, with no relative motion between any of them. Of course, for a reliable detection of such a dipole, the number of sources required may run into millions.

The NRAO VLA Sky Survey (NVSS), comprising 1.8 million discrete radio sources (Condon et al. 1998), is one such dataset that could be explored for signs of a radio dipole.  However, detailed calculations showed that a positive detection of such a dipole, at least to a  minimum acceptable significant level ($\stackrel{>}{_{\sim}} 3\sigma$), would indeed be not possible from a radio survey like the NVSS, as the statistical uncertainties would outweigh the signal indicated by the CMB dipole (Crawford 2009). It was envisaged that in order to detect a radio dipole at a statistically significant level, one might have to wait for, at least an order of magnitude larger, future radio surveys, like those planned from the Low Frequency Array (LOFAR) or the Square Kilometer Array (SKA) (Crawford 2009).
In fact, a couple of initial investigations of observed asymmetries in the number counts of distant radio sources in various directions in sky, because of the large known as well as unknown uncertainties, could not go beyond than to say that the radio source dipoles might be in concordance with the CMB dipole (Baleisis et al. 1998; Blake \& Wall 2002).
 
Nevertheless, in spite of these discouraging prognostications, a positive detection of a dipole at a statistically significant ($\sim 3\sigma$) level could be made from the NVSS data, in the number counts as well as in the sky brightness, which first time gave a rather high value of the solar peculiar velocity, $\sim 4$ times that expected from the CMB dipole (Singal 2011). In spite of the statistical uncertainties, which as predicted, were of the same order as the CMB dipole amplitude, a positive detection became possible only because the dipole amplitude had unexpectedly, turned out to be much larger than the CMB dipole. Another surprising finding was that in spite of the large difference in amplitudes, the direction of the NVSS dipole, coincided within statistical uncertainties, with the CMB dipole.

With the hope that it might help resolve the issue of the difference seen in amplitudes of the NVSS and CMB dipoles, another large radio catalogue, TIFR GMRT Sky Survey (TGSS), comprising 0.62 million sources (Intema et al. 2017), was investigated for any dipole asymmetries. However, it showed instead an even larger dipole anisotropy, amounting to a peculiar velocity $\sim 10$ times the CMB value, at a very significant level ($>10\sigma$) (Bengaly et al. 2018; Singal 2019a). Again, the TGSS dipole turned out to be in the same direction as the CMB dipole.
Many independent research groups have since reinvestigated the NVSS and/or TGSS catalogues, complementing with some other radio data and confirmed the radio dipoles to be of such larger amplitudes than the CMB dipole, though pointing broadly in the same direction (Rubart \& Schwarz 2013; Tiwari et al. 2015; Colin et al. 2017;  Siewert et al. 2021; Secrest et al. 2022; Wagenveld, Kl\"ockner \& Schwarz 2023).

Similar apparent anomalies in dipole amplitudes have been seen in sources selected from surveys other than the radio. For instance, active galactic nuclei (AGNs) picked from the Wide-field Infrared Survey Explorer (WISE) catalogue (Wright et al. 2010, Mainzer at al. 2014) too have yielded dipoles much larger than the CMB dipole (Singal 2021; Secrest et al. 2021,22; Dam, Lewis \& Brewer 2022; Kothari et al. 2022). 
In a~homogeneous selected DR12Q sample of quasars, drawn from the Sloan Digital Sky Survey (SDSS) III, a redshift dipole was seen, where an excess in quasar redshifts was observed in the direction of the CMB dipole. When interpreted in terms of the solar peculiar motion, the observed  excess in redshift distribution translated to a velocity $\sim 6.5$ times the CMB dipole in a direction directly opposite to, but~nonetheless parallel to, the~CMB dipole (Singal 2019b). 

Also a peculiar motion of the observer can introduce a dipole in the $m-z$ Hubble plot by affecting the observed redshifts $z$ as well as magnitudes $m$ of the sources, 
adversely in two opposite hemispheres. This has been exploited to estimate peculiar velocities from the Hubble diagrams of Supernovae Ia and of quasars with spectroscopic redshifts (Singal 2022a,b). Additionally, from Supernovae Ia, quasars and gamma-ray bursts, there is evidence for a higher Hubble constant in the CMB dipole direction (Krishnan et al. 2022; Luongo et al. 2022).

Many of the dipoles, radio as well as non-radio, seem to yield different amplitudes, though their directions seem to be pointing in a narrow region of sky around the CMB dipole, which has a rather low probability  ($\stackrel{\leq}{_{\sim}}10^{-5}$), to occur by a random chance (Singal 2023). A significant difference in amplitudes of various dipoles would imply a relative motion between the corresponding cosmic reference frames or the presence of inhomogeneities and anisotropies on cosmic scales, either of which will be against the CP on which the whole modern cosmology is based upon. The question whether the observable Universe is consistent with the CP has been reviewed in detail by Aluri et al. (2023). 

On the other hand, Mittal, Oayda \& Lewis (2024) from a Bayesian analysis of the Quaia sample of quasars found a dipole largely consistent with the CNB dipole.They find that the Quaia sample is influenced by selection effects with significant contamination near the Galactic Plane. After excising these regions, they find significant evidence that the Quaia quasar dipole is consistent with the CMB dipole, both in terms of the expected amplitude and direction. Their result seems to lend support to the cosmological principle. Additionally, Cheng et al. (2023) demonstrated that the difference in dipole amplitudes between NVSS and CMB can be fully accounted for by incorporating clustering and shot-noise contributions to the total NVSS dipole and conclude that the NVSS dipole is consistent with a kinematic origin for the CMB dipole within $\Lambda$CDM.

More recently, two other large radio surveys, VLASS (Gordon et al. 2021) and RACS (Hale et al. 2021), have shown dipole asymmetries in their number counts to be not only of significantly different amplitudes, even their directions do not seem to be consistent with each other (Singal 2023). An initial claim that a dipole derived from the combined VLASS and RACS data is consistent in both magnitude and direction with the CMB dipole (Darling 2022), was soon shown to be following a procedure of combining data from two independent surveys with partial sky coverages which could lead to false results. On the other hand, the dipoles determined separately from the two surveys, are congruent neither with each other nor with other such dipoles from previous large radio surveys, like the NVSS and TGSS, and certainly not with the CMB dipole (Singal 2023).

The four radio surveys,  NVSS, TGSS, VLASS and RACS, have large overlaps in the sky coverage and, accordingly, should have a large fraction of radio sources common to all of them. Therefore significant differences between their estimated radio dipoles should be disconcerting all the more. 
It is, therefore, imperative that a detailed investigation of the incongruency observed in dipoles, estimated from different radio source surveys, be carried out in order to, hopefully, ascertain what could mostly be the cause of this incongruency and if possible, to mitigate its effects up to a sufficient level to help resolve the incongruency to arrive at a consistent picture for the radio dipole. 

\section{Dipole from number counts in a radio survey with a partial sky coverage}\label{S2}
According to the CP, the number density of radio sources per unit solid angle seen by a comoving observer, that is an observer stationary in the expanding cosmic fluid, should be independent of the direction in sky, irrespective of the  number density or luminosity evolutions with redshift of the underlying AGN population. Let $N_0$ be such a number density, per unit solid angle, for the class of sources under investigation. However, if due to whatever reason the number density is not uniform across the sky and there is, say, a dipole in the number distribution, then we can express the number density as a function of $\theta$, the polar angle with respect to the direction of the dipole in sky, as
\begin{equation}
\label{eq:1}
 N(\theta) 
 =  N_0(1+{\cal D}\cos \theta)\,.
\end{equation}
Amplitude, ${\cal D}$, of the dipole could be determined from an integral over solid angle, $\Omega$, of the sky 
\begin{eqnarray}
\label{eq:2}
 \frac {\int N(\theta)\cos \theta\:\mathrm{d}\Omega}{\int N(\theta)|\cos \theta|\:{\rm d}\Omega}&=&
\int_{0}^{\pi}{\cal D}\cos^2 \theta \sin\theta\: \mathrm{d}\theta= \frac {2{\cal D}}{3}\,.
\end{eqnarray}

If our radio source survey has a complete-sky coverage, then the dipole direction in sky can be estimated from the vector sum, $\sum_{i=1}^{N}\bf{\hat{r}}_{\rm i}$, where $\bf{\hat{r}}_{\rm i}$ is the angular position vector of $i$th source in the survey data (Crawford 2009; Singal 2011). The integral of number density over the sky in Eq.~(\ref{eq:2}) can be converted  into a normalized sum of the position vector components along the direction of the dipole for all sources in the sample to write the dipole amplitude as 
\begin{eqnarray}
\label{eq:3}
{\cal D}=\frac {3}{2}\frac{\Sigma \cos \theta_i}{\Sigma |\cos \theta_i|}\,,
\end{eqnarray}
where $\theta_{\rm i}$ is the polar angle of the $i${th} source with respect to the direction of the dipole. The error in the value of computed dipole amplitude, ${\cal D}$, is $\sqrt {3/N}$, where $N$ is the total number of sources in the sample (Crawford 2009; Singal 2011). 
This method to determine the dipole is known as the vector method. 

However, dipole amplitude $\cal D$ can be estimated in another way, known as the hemisphere method, although, unlike the vector method, it does not directly yield the dipole direction in sky, One starts with an assumed pole in sky, then dividing the sky in two equal hemispheres, $H_1$ and $H_2$, with $H_1$ centred on the pole direction and $H_2$ centred on the anti-pole direction, we can count in our sample the number of sources within each hemisphere. Let $N_1,N_2$ be the number of sources in $H_1, H_2$ respectively. Then we can write  
\begin{eqnarray}
\label{eq:4}
\frac {(N_1 - N_2)} {N_1 + N_2}&=& \frac {{\int_{H_1}N(\theta) \:\mathrm{d}\Omega}-{\int_{H_2} N(\theta) \:\mathrm{d}\Omega}}{{\int_{H_1}N(\theta) \:\mathrm{d}\Omega}+{\int_{H_2} N(\theta) \:\mathrm{d}\Omega}}\nonumber\\
&=&\int_{0}^{\pi/2} {\cal D}\cos \theta \:\sin\theta\: \mathrm{d}\theta= \frac {{\cal D}}{2}\,.
\end{eqnarray}
The error in  ${\cal D}$ thus determined is $\sqrt {2/N}$, where $N=N_1 + N_2$ is the total number of sources in the sample. 

Employing the hemisphere method, one can apply `brute force' to determine an optimum value for the dipole position in sky (Singal 2019b,23). Briefly, first whole $4\pi$ sr of sky is divided into $m$ pixels of $\Delta\psi\times \Delta\psi$ solid angle each. Then taking the centre of each of these $m$ cells to be the pole, the dipole magnitude $p_{\rm i}$  for $i=1$ to $m$ is computed, using the hemisphere method (Eq.~(\ref{eq:4})). This, however, yields only a projection of the dipole in the direction of $i$th pixel, which should have a $\cos\psi_{\rm i}$ dependence where $\psi_{\rm i}$ is a polar angle of $i$th pixel with respect to the, as yet unknown, actual pole. 
Then taking each of the $m$ positions in turn as pole, a 3-d $\cos\psi$ fit to the $p_{\rm i}$ values of its neighbouring $m-1$ pixels is made, and a chi-square value for each of these $m$ fits is computed, whose minimum yields the optimum location of the dipole in sky. One could then compare this dipole, both position and the corresponding magnitude, with that obtained from the vector method. This match could be important, especially in case the sky is only partially covered in the survey sample. 

In general a radio survey may not cover the whole sky, basically due to the geographical location of the radio telescope, which may be away from the equator and thus a section of the northern or southern hemisphere of the sky may not be visible to it.  For instance the NVSS survey, conducted by the Very Large Array, which is geographically located at a latitude $+34^{\circ}$ cannot see the whole southern sky and thus covers the sky north of declination $-40^{\circ}$ (Condon et al. 1998). In such a  case ${\cal D}$ can be determined by restricting the source sample to, say, $-40^{\circ}$ to $+40^{\circ}$ in declination (Singal 2011). 
Moreover, we may also have a zone of avoidance about the galactic plane. Exclusion of such sky-strips, which affect the forward and backward measurements 
identically, for example symmetric strips in diametrically opposite regions on the sky, do not have effects on the direction inferred for the dipole (Ellis \& Baldwin 1984, Singal 2011),
except that errors in estimated position may become larger because of lesser data. 

However, Eqs.~(\ref{eq:3}) and (\ref{eq:4}) for ${\cal D}$ are valid for a full coverage of the sky, therefore amplitudes of the dipole thus determined could be affected by a factor of the order of unity, depending upon the gaps in sky coverage. We could then modify Eqs.~(\ref{eq:3}) and (\ref{eq:4}) as 
\begin{eqnarray}
\label{eq:5}
{\cal D}=\frac {3}{2 k_1}\frac{\Sigma \cos \theta_i}{\Sigma |\cos \theta_i|}\,,
\end{eqnarray}
and 
\begin{equation}
\label{eq:6}
{\cal D}=\frac {2}{k_2}\:\frac {(N_1 - N_2)} {N_1 + N_2}\,..
\end{equation}
where $k_1,k_2$ are constants, of the order of unity, to be determined numerically for the effect of gaps.

\begin{table}
\begin{center}
\caption{\label{T1}Amplitude correction factors $k_1$ and $k_2$ for radio dipoles in the direction of CMB dipole  determined using data having a limited sky coverage.}
\hskip4pc\vbox{\columnwidth=33pc
\begin{tabular}{lccccccccccccccc}
\hline 
(1)&(2)&(3)\\
Sky coverage limit & $k_1$ &  $k_2$ \\
 & Vector dipole & Hemisphere method \\ 
\hline
$|{\delta}|\leq90^\circ,|b|>00^\circ$ & 1.00 & 1.00 \\
$|{\delta}|\leq90^\circ,|b|>10^\circ$ & 1.04 & 1.03 \\
$|{\delta}|\leq90^\circ,|b|>20^\circ$ & 1.07 & 1.08 \\
$|{\delta}|\leq90^\circ,|b|>30^\circ$ & 1.10 & 1.14 \\\\
$|{\delta}|\leq53^\circ,|b|>10^\circ$ & 1.08  &1.17 \\
$|{\delta}|\leq40^\circ,|b|>10^\circ$ & 1.13  &1.24 \\
$|{\delta}|\leq30^\circ,|b|>10^\circ$ & 1.16  &1.28  \\\\
$|{\delta}|\leq25^\circ,|b|>10^\circ$ & 1.18 & 1.29 \\
$|{\delta}|\leq20^\circ,|b|>10^\circ$ & 1.19  &1.30 \\
$|{\delta}|\leq15^\circ,|b|>10^\circ$ & 1.20 & 1.31 \\
\hline
\end{tabular}
}
\end{center}
\end{table}
Table 1 shows values of correction factors $k_1$ and $k_2$ for various sky coverage limits, which may also have an appropriate zone of avoidance about the galactic plane, computed for a radio dipole that may lie along the CMB dipole direction. For other dipole directions, values  of correction factors may differ somewhat and would thus  need to be computed numerically in  individual cases. We have calculated $k_1$ and $k_2$ in each case with respect to the direction actually measured for that case, which might be different from the CMB dipole direction. It should be noted that for no restriction limits on the declination ($\delta \leq90^\circ$) as well as on the galactic latitude ($|b|>0$), there may be no correction factors ($k_1=k_2=1$), however for finite $|b|$ limits, the correction factors may still differ from unity even if the sky otherwise were fully covered in the survey.

Dipole ${\cal D}$ comes directly from the actual observational data, while its  origin might be a matter of interpretation. Its genesis, in conventional wisdom, lies in a peculiar motion of the Solar system, where the observer along with the Solar system is moving with a peculiar velocity $\bf v$, then the stellar aberration and Doppler boosting would give rise to a dipole in the number density, along the direction of peculiar motion, with an amplitude  
\begin{equation}
\label{eq:7}
{\cal D}=[2+ x (1+\alpha)]\frac {v}{c}\,.
\end{equation}
where $c$ is the velocity of light, $\alpha$ ($\approx 0.8$) is the spectral index, defined by $S \propto \nu^{-\alpha}$, and $x$ is the index of the integral source counts of extragalactic radio source population, which follows a power law, $N_0(>S)\propto S^{-x}$ ($x \sim 1$) (Ellis \& Baldwin 1984). 

If the dipole represents indeed a genuine Solar peculiar motion, then its value should not depend upon the sample or the technique used to determine it. In particular, peculiar velocity estimates from different samples of distant radio sources should all be the same as the value inferred from the CMB. For convenience of comparison we use a parameter $p$ for the amplitude of the peculiar velocity $v$, in units of the CMB value 370 km s$^{-1}$, so that $v=p \times 370$ km s$^{-1}$, with $p=0$ implying a nil peculiar velocity while $p=1$ meaning the CMB value.

\section{Selection of identical sky coverage for each of four radio surveys for dipoles determination}\label{S3}
Each of the four radio surveys, NVSS, TGSS, VLASS and RACS, spans a large fraction of the total sky contiguously, except may be for a zone of avoidance about the Galactic plane. The NVSS catalogue covers sky north of declination $-40^{\circ}$, thus covering 82\% of the whole sky and contains $\sim 1.8$ million sources with a flux-density limit $S>3$ mJy  at 1.4 GHz, carried out with the Very Large Array (Condon et al. 1998). The TIFR GMRT Sky Survey (TGSS) carried out at 150 MHz with the Giant Metre-wave Radio Telescope (Swarup et al. 1991), covers sky north of declination $-53^{\circ}$, covering 90\% of the whole sky, comprises 0.62 million sources (Intema et al. 2017).  
The VLASS catalogue at 3 GHz (Gordon et al. 2021), carried out at Karl G. Jansky Very Large Array (Lacy et al. 2020), covers the sky north of $-40^\circ$ declination, thus covering 82\% of the whole sky and contains 1.9 million radio sources, The RACS catalogue carried out using the Australian Square Kilometre Array Pathfinder (ASKAP), will ultimately cover the sky south of $+51^\circ$ declination (McConnell et al. 2020), though the presently available data (Hale et al. 2021) cover the sky south of $+30^\circ$ declination, which is 75\% of the total sky, containing 2.1 million radio sources at 887.5 MHz, 
\begin{figure}
\includegraphics[width=\columnwidth]{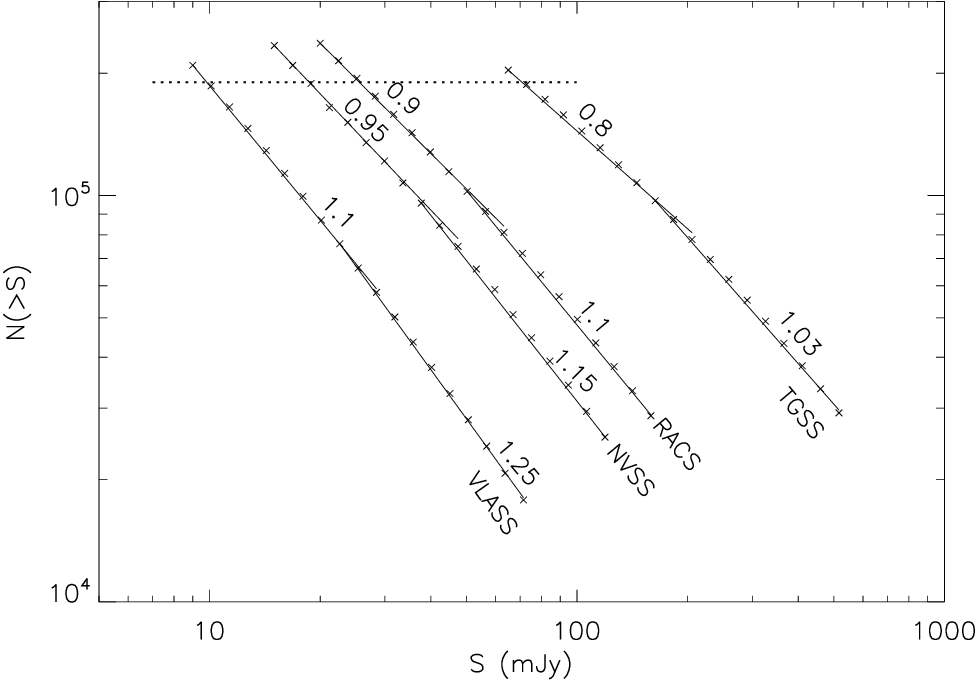}
\caption{A plot of the integrated source counts $N(>S)$ against $S$, for the four radio  samples, showing the power law  behavior ($N(>S)\propto S^{-x}$) of the source counts. From  piece-wise straight line fits to data in different flux-density ranges in each sample, index $x$ appears to steepen for stronger sources, as shown by continuous lines with the best-fit $x$ values written above. The horizontal dotted line, representing a constant  number of sources, helps fixing flux-density levels for various surveys so that there are about the same number of sources in each survey and thereby comprising essentially the same set of sources.}
\label{F1}
\end{figure}
\begin{figure}
\includegraphics[width=\columnwidth]{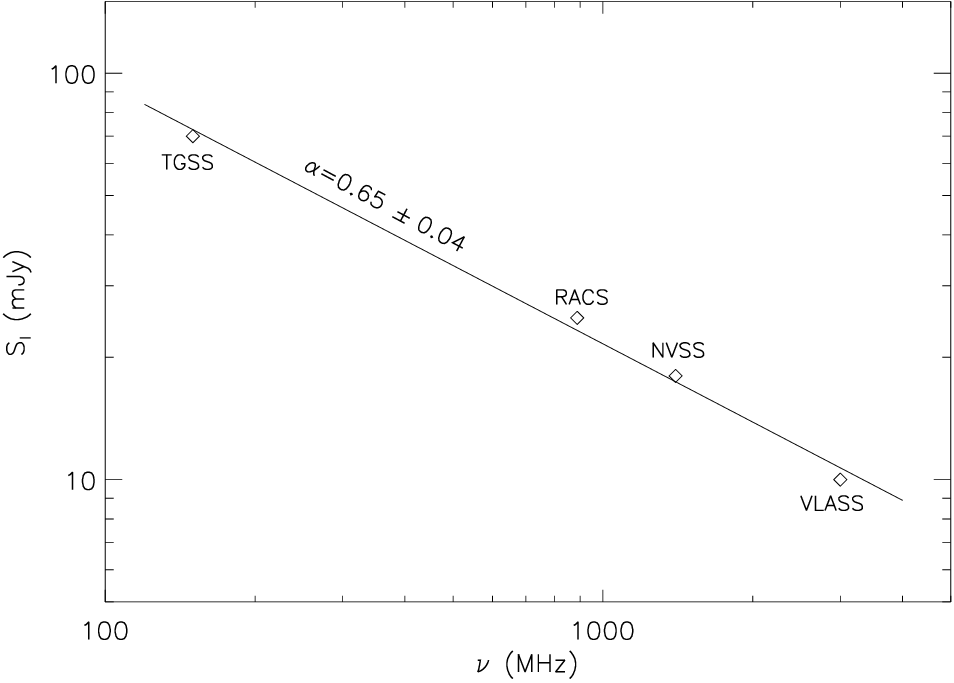}
\caption{Average value of spectral index determined by a least square fit to the flux-density  $S_l$ of different surveys, which are at different frequencies, by choosing the same source count level from the integral source count plots in Fig.~\ref{F1}.} 
\label{F2}
\end{figure}

Thus these surveys have large overlaps in the sky coverage and a major fraction of the total sky is common to all the four radio catalogues, which could be examined in details to understand the occurrence of different dipole asymmetries seen in these four individual catalogues. 
Accordingly if we choose a big enough common sky region for all four surveys by introducing symmetric cuts in declination, and determine dipoles in all four cases from the data limited to that common sky region (say, within a declination range $|\delta| < 30^\circ$), then we expect thus determined dipoles to be congruent, irrespective of their underlying source of origin. After all any problems related to a limited sky coverage (like declination within $\pm 30^\circ$)  or even galactic plane gap ($|b|\leq 10^\circ$), would be common for all the four cases. 
\begin{table*}
\begin{center}
\caption{\label{T2}Estimates of peculiar velocity from the same sky coverage ( $0^\circ<|{\delta}|\leq30^\circ$, $|b|>10^\circ$) for various radio source samples.}
\begin{tabular}{ccccccccc}
\hline 
(1)&(2)&(3)&(4)&(5)&(6)&(7)&(8)\\
Serial& Sample& $\nu$ &Flux-density &  $N$ 
&\multicolumn{3}{c}{Peculiar velocity vector estimate} \\
\cline{6-8}
No.&& &$S$& &  RA & Dec &   $v$\\
&& (GHz)&(mJy) && ($^{\circ}$)&  ($^{\circ}$) &   ($10^3$ km s$^{-1}$) \\ 
\hline
1 & VLASS& 3.0& $S\geq10$ &  188021 & $212\pm 13$ & $+21\pm 16$ & $1.1 \pm 0.3$\\ 
2 & NVSS& 1.4&$S\geq18$ &  198230 & $163\pm 12$ & $+05\pm 10$ &  $1.1 \pm 0.3$\\ 
3 & RACS& 0.9& $S\geq25$ &  195427 &  $196\pm 14$ &  $-41\pm 25$ &$3.2\pm 0.4$\\  
4 & TGSS& 0.15& $S\geq70$ &  193280 & $138\pm 15$ & $-01\pm 12$ & $3.3\pm 0.4$\\ 
5 & CMB &&&& 168& $-07$ & 0.37\\
\hline
\end{tabular}
\end{center}
\end{table*}

In Fig.~\ref{F1} we have plotted the integrated source counts $N(>S)$ for different $S$ in the common coverage area of the sky region of $|\delta| < 30^{\circ}$ and $|b|> 10^\circ$, for all four radio samples. The index $x$  in the power law relation, $N(>S)\propto S^{-x}$, can be estimated from the slope of the $\log N-\log S$ plot in Fig.~\ref{F1}, where we find that the index $x$ slightly steepens from low to high flux-density levels for all samples. In order to ensure that the sources used in different radio surveys are mostly common, we have chosen the flux-density levels of individual surveys such that they have approximately the same number of sources in our sub-samples. Intersection of the horizontal dotted line in Fig.~\ref{F1} at around $N(>S_l)\sim 1.9 \times 10^5$ indicates 
flux-density levels for various surveys to be $S_l=10$ mJy for the VLASS, $S_l=18$ mJy for the NVSS, $S_l=25$ mJy for the RACS and $S_l=70$ mJy for the TGSS. Choosing these flux-density limits for various surveys ensures that in different surveys we are employing approximately the same set of sources and would therefore expect to encounter similar dipole asymmetries. This is because with the sky coverage exactly matching, most sources should be common in all the four surveys, so one should expect the dipoles to be similar in all the four surveys, assuming no systematics in various data sets.

Actually for a constant spectral index $\alpha$, all $N$ sources in the integrated counts at a flux-density greater than $S_1$, in a survey made at a frequency $\nu_1$, will be present in the integrated counts of another survey made at a frequency $\nu_2$ at a flux-density greater than $S_2=S_1 (\nu_1/\nu_2)^\alpha$. In reverse, for two surveys made at $\nu_1$ and  $\nu_2$, by determining respectively $S_1$ and  $S_2$ for a fixed integrated source count number $N$ and assuming that the spectral index is relatively constant, one could determine $\alpha$ from the slope in the ($\ln S-\ln\nu$) plot. Such estimates of $S_1$, $S_2$ etc. are quite robust, since differential number counts are given by ${\rm d}N/{\rm d}S\propto S^{-x-1}\sim S^{-2}$, thus sources in any chosen flux-density range bunch together near the lower flux-density level. Figure~\ref{F2} shows the average value of spectral index determined by a least square fit to the flux-density of different surveys, which are at different frequencies, by choosing the same source count level from the integral source count plots, as shown by a horizontal dotted line in Fig.~\ref{F1}. The value $\alpha=0.65$ thus obtained, which we shall be using in Eq.~(\ref{eq:7}) to compute the peculiar velocity $v$ from the dipole amplitude $\cal D$, is slightly less than the normally assumed value $\alpha=0.8$, however for $x\sim 1$, it makes a difference of only $\sim 4$ percent in the computed $v$ values, which is insignificant compared to the factor of $\sim 4$ by which, as we shall see later, the peculiar velocity estimates from radio surveys appear higher than those inferred from the CMB maps.

Table~\ref{T2} shows the peculiar velocities derived in all four radio surveys, with data within  $|\delta| \leq 30^\circ$ and $|b|> 10^\circ$. We have determined the dipoles both from the vector method as well as the hemisphere method, where amplitudes were determined from Eqs.~(\ref{eq:5}) or (\ref{eq:6}), with the application of $k_1$ and  $k_2$ corrections for the gaps in the sky coverage, computed  for corresponding dipole direction, which gave reasonably consistent values for the two dipole amplitudes. 
In the hemisphere method, we used the brute force method (Section \ref{S2}) with $\Delta\psi=10^\circ$ and $m=422$, to determine the optimum dipole position. This dipole position in each sample turned out, perhaps not surprising, to be almost identical to that determined from the vector method, which is entered in Table~\ref{T2}.

The values tabulated for the dipole amplitudes in Table~\ref{T2}, as well as later in Tables~\ref{T3} to \ref{T7} (Section \ref{S4}), are the weighted means of the two values determined from Eqs.~(\ref{eq:5}) and (\ref{eq:6}). If the two individual values were statistically independent, the final errors in the weighted means would be less on the average by a factor of $\sqrt{2}$.  However, the two individual values may not be  statistically independent, therefore we have multiplied the computed error of the  weighted mean by a factor of  $\sqrt{2}$. 

The estimated dipoles for different surveys show large differences, both in the directions and amplitudes, given in rows one to four of Table~\ref{T2}, where we have also listed the CMB dipole position (row five), which, in comparison, has negligible position uncertainties and has a peculiar velocity, $v=370$ km s$^{-1}$ along RA$=168^{\circ}$, Dec$=-7^{\circ}$. In all cases the amplitude of the peculiar velocity for the radio source dipole appears much higher than that estimated for the CMB dipole.  We notice that while the total number $N$ of sources at each flux-density level do reasonably well match within a few percent, yet not only the direction estimated for the velocity vectors vary from one radio survey to another, even derived magnitudes of dipole $\cal D$ and the peculiar velocity $p$ therefrom, differ as much as by a factor of $\sim 3$ between various radio surveys. Of course all estimates of the peculiar velocity $v$ in all radio data are way above the values expected from the CMB (by as much as a factor of $\sim 3$ to $\sim 9$), while, from the CP, these dipoles, both in the direction and magnitude, should be concordant with the CMB dipole.
\begin{figure*}
\includegraphics[width=14cm]{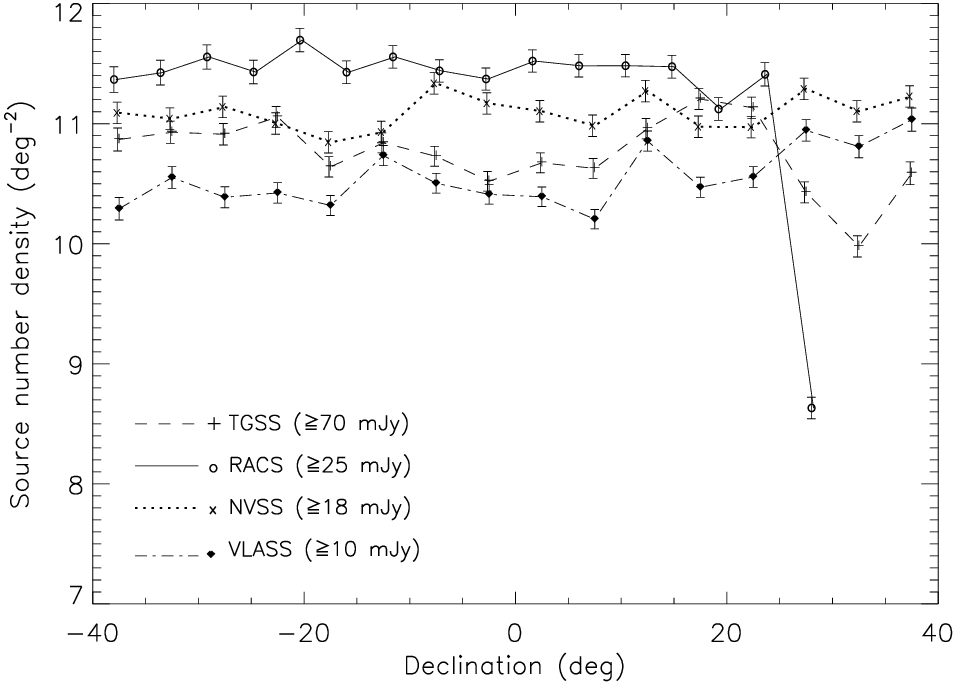}
\caption{Number density of sources with declination in steps of $5^\circ$ for the four radio surveys. 
\label{F3}}.
\end{figure*}

In Table~\ref{T2}, there seem no appreciable changes in the dipole direction or amplitude in the  NVSS case when going from $|\delta| \leq  40^\circ$ (Singal 2011) to $|\delta| \leq 30^\circ$ or in the TGSS case when going from $|\delta|\leq 53^\circ$ (Singal 2019a) to $|\delta| \leq 30^\circ$ limits. Moreover, no changes can be in  seen in the RACS case from Singal (2023), as the declination limits remains unchanged at $|\delta| \leq 30^\circ$. However, for the VLASS data, when the declination limits are changed from $|\delta| \leq 40^\circ$ (Singal 2023) to $|\delta| \leq 30^\circ$ (Table~\ref{T2} here), even though the dipole amplitude remains almost unchanged, the derived dipole position seems to change in declination from $+42^\circ$ to $+21^\circ$. This could perhaps be an indication of some declination-dependent systematics in the survey data whose effect we strive to possibly mitigate here in the dipole estimates. Further, as has been pointed out in Singal (2023), it appears that any declination-dependent systematics in the VLASS data seem to become more pronounced for weaker sources, i.e. with a decreasing flux-density of the sample. If there were no systematics in the source count data,  then the dipole direction should not shift systematically in a catalogue with changing dec limits apart from any possible increase in errors. No such shifts, for example, were seen in the NVSS or TGSS data.
\section{Dipoles determined from sky coverage increasingly away from the declination limit for each survey}\label{S4}
\begin{figure*}
\includegraphics[width=14cm]{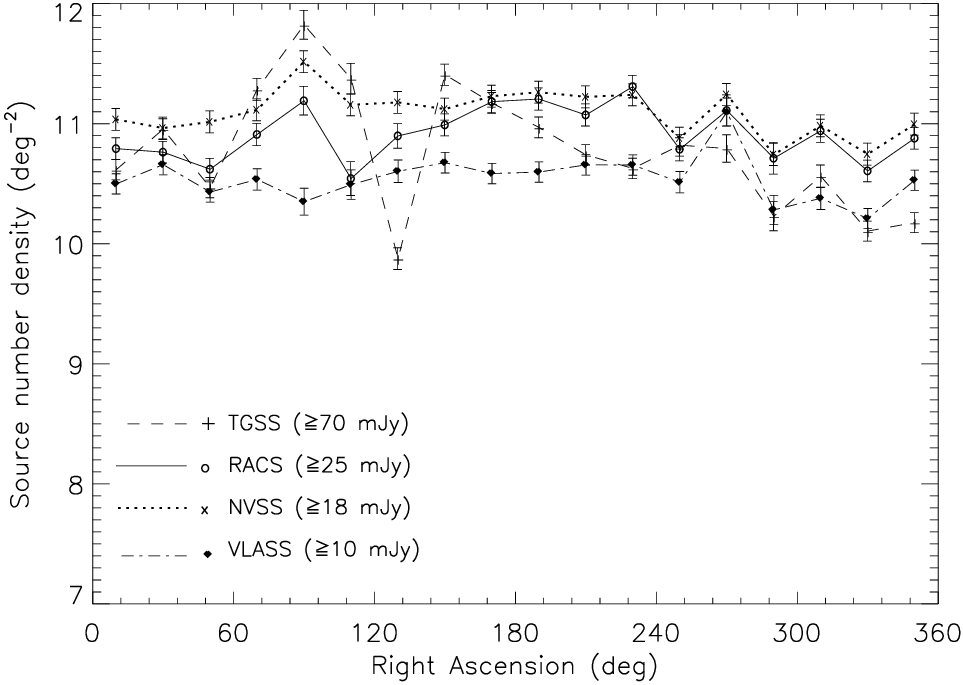}
\caption{Number density of sources as a function of right ascension in steps of $20^\circ$ for the four radio surveys. 
\label{F4}}.
\end{figure*}

In order to ascertain if there are tractable large scale systematics in various radio surveys, we have plotted in Figs.~\ref{F3} and \ref{F4} radio source densities as a function of declination (in steps of $5^\circ$) and right ascension (in steps of $20^\circ$) for all four surveys. From density plots for the VLASS and the RACS data, Darling (2022) had noted that the source density in the VLASS data dropped by about $16\%$ from declination $-15^\circ$ to $-40^\circ$, while for the RACS data a $6\%$ decrease was seen for declinations $> +25^\circ$; neither survey showed any such trends in right ascension.  We also find a sharp drop in the number density in the RACS sample near the northern declination limit of the  RACS data, much more than pointed out by Darling (2022). At the same time, we do not see any density drops of the sort noted by Darling (2022) in the VLASS sample. Also no systematic variations in the source densities for the NVSS or TGSS samples are seen with declination. Further, we also find no systematic variations in the number densities as a function of right ascension in any of our four samples (Fig~\ref{F4}).

\begin{table*}
\begin{center}
\caption{\label{T3}Velocity vector from number counts for the RACS dataset with $|b|>10^\circ$.}
\hskip4pc\vbox{\columnwidth=33pc
\begin{tabular}{cccccccccccccccc}
\hline 
(1)&(2)&(3)&(4)&(5)&(6)&(7)&(8)\\
Serial&Flux-density & Declination  & $N$ 
&\multicolumn{4}{c}{Peculiar velocity vector estimate} \\
\cline{5-8}
No.&$S$ &Coverage & &  RA & Dec & ${\cal D}$  &   $p$\\
&(mJy)&($^{\circ}$)&& ($^{\circ}$)&  ($^{\circ}$) &  ($10^{-2}$) & ($370$ km s$^{-1}$)&  \\
\hline
1 & $S\geq 60$ &$|{\delta}|\leq30$ &  85707 &  $198\pm 17$ &  $-43\pm 30$ & $4.1\pm 0.7$ & $8.4\pm1.4$ \\
2 & $S\geq 60$ &$|{\delta}|\leq29$ &  84264 &  $197\pm 17$ &  $-13\pm 08$ & $2.1\pm 0.5$ & $4.3\pm1.1$\\
3 & $S\geq 60$ &$|{\delta}|\leq25$ &  73583 &  $203\pm 18$ &  $-10\pm 08$ & $2.2\pm 0.6$ & $4.4\pm1.2$ \\
4 & $S\geq 60$ &$|{\delta}|\leq20$ &  59723 &  $196\pm 19$ &  $-01\pm 08$ & $2.1\pm 0.6$ & $4.2\pm1.2$ \\
5 & $S\geq 60$ & $|{\delta}|\leq15$ & 45280 &  $195\pm 19$ &  $+03\pm 08$ & $1.7\pm 0.7$ & $3.5\pm1.4$ \\\\
\\6 & $S\geq 40$ & $|{\delta}|\leq30$ & 127760 &  $195\pm 14$ &  $-42\pm 28$ & $3.9\pm 0.6$ & $9.2\pm1.3$ \\
7 & $S\geq 40$ & $|{\delta}|\leq29$ & 125577 &  $194\pm 14$ &  $-11\pm 12$ & $2.0\pm 0.4$ & $4.6\pm1.0$ \\
8 & $S\geq 40$ & $|{\delta}|\leq25$ & 109534 &  $194\pm 15$ &  $-10\pm 12$ & $2.0\pm 0.5$ & $4.7\pm1.1$ \\
9 & $S\geq 40$ & $|{\delta}|\leq20$ & 88919 &  $191\pm 15$ &  $-05\pm 13$ & $1.8\pm 0.5$ & $4.2\pm1.2$ \\
10 & $S\geq 40$ & $|{\delta}|\leq15$ & 67477 &  $194\pm 16$ &  $+03\pm 13$ & $1.5\pm 0.6$ & $3.5\pm1.3$ \\\\
\\11 & $S_l\geq 25$ &$|{\delta}|\leq30$ &  195427 &  $196\pm 14$ &  $-41\pm 25$ & $3.7\pm 0.4$ & $8.6\pm 1.0$ \\
12 & $S_l\geq 25$ &$|{\delta}|\leq29$ &  192040 &  $196\pm 16$ &  $-10\pm 11$ & $2.0\pm 0.4$ & $4.7\pm 0.8$ \\
13 & $S_l\geq 25$ &$|{\delta}|\leq25$ &  167721 &  $194\pm 15$ &  $-09\pm 11$ & $2.0\pm 0.4$ & $4.7\pm 0.9$ \\
14 & $S_l\geq 25$ &$|{\delta}|\leq20$ &  136102 &  $194\pm 14$ &  $-04\pm 12$ & $1.9\pm 0.4$ & $4.5\pm 0.9$ \\
15 & $S_l\geq 25$ &$|{\delta}|\leq15$ &  103468 &  $193\pm 16$ &  $+02\pm 14$ & $1.6\pm 0.5$ & $3.6\pm 1.1$ \\
\hline
\end{tabular}
}
\end{center}
\end{table*}
Although some large systematics might hopefully be obvious in the density plots as seen, e.g. in the RACS data, it should nevertheless be noted that an absence of noticeable trend against right ascension or/and declination is no guarantee that no systematics in the data are present that could affect the dipole estimate. After all it is only a density difference of the order of one percent (less than the size of error bars in Figs.~\ref{F3} and \ref{F4}) between the pole and anti-pole directions that could give rise to an artificial dipole larger than the CMB dipole. 
\begin{figure*}
\includegraphics[width=\linewidth]{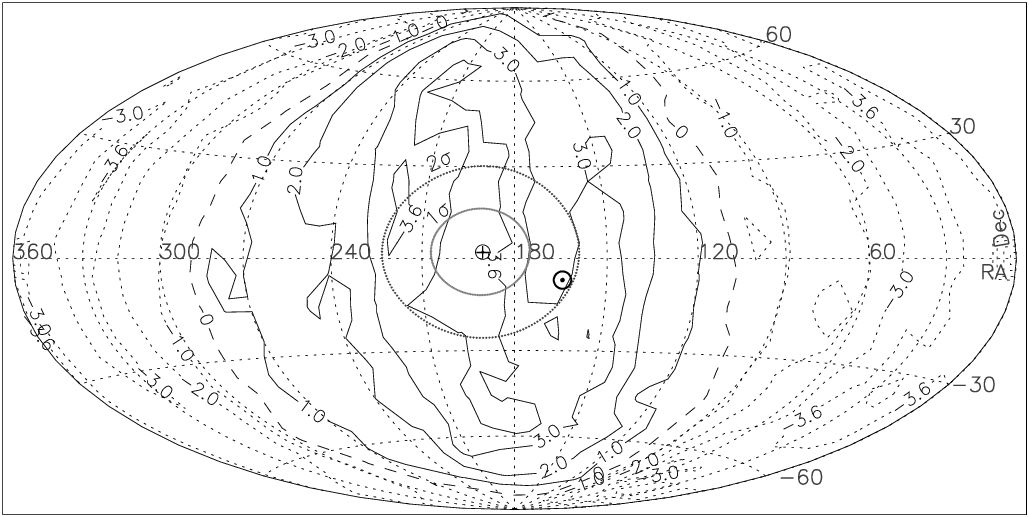}
\caption{A contour map of the dipole amplitudes, estimated for various directions in the sky, from the RACS data restricted to $|\delta| \leq 15^\circ$ and $S_{\rm l}\geq 25$ mJy. The~horizontal and vertical axes denote RA and Dec in degrees. The~true pole direction is expected to be closer to the higher contours values, shown by continuous lines, while the negative contour values are shown by dotted lines. The~dashed curve represents the zero amplitude of the dipole. 
The symbol $\oplus$ indicates the derived pole position for the RACS sample, along with   $1\sigma$ and $2\sigma$ error ellipses around it. The symbol $\odot$ indicates the CMB pole position. 
\label{F5}
} 
\end{figure*}

One simple and direct way to avoid declination-dependent systematics near the declination-coverage limits of the survey is to restrict the data to be used in the dipole estimate to a declination range away from the declination limit of the survey. This is hinted by the VLASS data when we restricted the sample range to $|\delta|\leq 30^\circ$ from the earlier range of $|\delta|\leq 40^\circ$ used by Singal (2023), the derived dipole declination  moved from $+42^\circ$ to  $+21^\circ$, closer toward the estimates from other radio surveys like the NVSS and TGSS that gave dipole direction consistent with the CMB dipole. Taking a cue from this we have tried declination range reduced in steps of $5^\circ$, moving away from the declination limits of each survey. Of course, one has to stop at some minimum value of the declination range, since errors increase as the number of sources in the sub-sample reduce with decreasing sky coverage, We find that the minimum useful sky coverage that gives still useful results for the dipole is $|\delta|\leq 15^\circ$. With $|\delta|\leq 30^\circ$, the coverage is 50 percent of the total sky area, while $|\delta|\leq 15^\circ$ reduces it to 25 percent. Of course there is an additional limit of zone of avoidance about the galactic plane ($|b|< 10^\circ$). As mentioned earlier, exclusion of symmetric strips in diametrically opposite regions on the sky, for example, dropping symmetric bands of declination in northern and southern hemispheres, do not have systematic effects on the direction inferred for the dipole (Ellis \& Baldwin 1984, Singal 2011), Any effects on the dipole amplitude can be taken care of by employing corrections $K_1$ and $K_2$ as in Eqs.~(\ref{eq:5}) and (\ref{eq:6}).

There is a sharp drop in the number density of radio sources near the declination limit ($\delta\sim 30^\circ$) of the RACS survey (Fig.~\ref{F3}).  
Such a sharp change in source number density could not be due the presence of a dipole itself, which should be a smooth $\cos\theta$ pattern. Even otherwise a dipole like the CMB dipole require maximum change in the number densities in sky (between the pole and anti-pole directions) to occur only by less than one percent or so.
Now any such abrupt change in the source density, which may not be due to a genuine number counts in sky, could give rise to large shift in the dipole estimate, both in direction as well as magnitude.
\begin{table*}
\begin{center}
\caption{\label{T4}Velocity vector from number counts for the VLASS dataset with $|b|>10^\circ$.}
\hskip4pc\vbox{\columnwidth=33pc
\begin{tabular}{ccccccccccccccc}
\hline 
(1)&(2)&(3)&(4)&(5)&(6)&(7)&(8)\\
Serial&Flux-density & Declination  & $N$ 
&\multicolumn{4}{c}{Peculiar velocity vector estimate} \\
\cline{5-8}
No.&$S$ &Coverage & &  RA & Dec & ${\cal D}$  &   $p$\\
&(mJy)&($^{\circ}$)&& ($^{\circ}$)&  ($^{\circ}$) &  ($10^{-2}$) & ($370$ km s$^{-1}$)&  \\
\hline
1 & $S\geq 30$ & $|{\delta}|\leq40$ & 69465 &  $174\pm 15$ &  $+13\pm 25$ & $1.7\pm 0.6$ & $3.7\pm1.3$ \\
2 & $S\geq 30$ & $|{\delta}|\leq30$ & 54275 &  $195\pm 16$ &  $-02\pm 12$ & $1.9\pm 0.7$ & $4.0\pm1.4$ \\
3 & $S\geq 30$ & $|{\delta}|\leq25$ & 45682 &  $211\pm 16$ &  $-08\pm 13$ & $2.5\pm 0.7$ & $5.4\pm1.5$ \\
4 & $S\geq 30$ & $|{\delta}|\leq20$ & 37083 &  $214\pm 17$ &  $-05\pm 13$ & $2.8\pm 0.8$ & $6.0\pm1.7$ \\
5 & $S\geq 30$ & $|{\delta}|\leq15$ & 28288 &  $211\pm 19$ &  $-10\pm 14$ & $3.5\pm 0.9$ & $7.3\pm1.9$\\\\
\\6 & $S\geq 20$ & $|{\delta}|\leq40$ & 112322 &  $188\pm 14$ &  $+28\pm 23$ & $1.9\pm 0.5$ & $4.0\pm1.1$ \\
7 & $S\geq 20$ & $|{\delta}|\leq30$ & 87806 &  $210\pm 15$ &  $+03\pm 12$ & $1.9\pm 0.5$ & $3.9\pm1.1$ \\
8 & $S\geq 20$ & $|{\delta}|\leq25$ & 74117 &  $222\pm 15$ &  $-05\pm 13$ & $2.2\pm 0.6$ & $4.6\pm1.2$ \\
9 & $S\geq 20$ & $|{\delta}|\leq20$ & 60214 &  $222\pm 16$ &  $+00\pm 15$ & $2.3\pm 0.6$ & $5.0\pm1.4$ \\
10 & $S\geq 20$ & $|{\delta}|\leq15$ & 45775 &  $218\pm 17$ &  $-03\pm 14$ & $3.0\pm 0.7$ & $6.3\pm1.5$ \\\\
\\11 & $S_l\geq 10$ & $|{\delta}|\leq40$ & 240458 &  $189\pm 13$ &  $+42\pm 16$ & $2.0\pm 0.4$ & $4.2\pm 0.8$ \\
12 & $S_l\geq 10$ & $|{\delta}|\leq30$ & 188021 &  $212\pm 13$ &  $+21\pm 16$ & $1.4\pm 0.4$ & $3.1\pm 0.8$ \\
13 & $S_l\geq 10$ & $|{\delta}|\leq25$ & 158986 &  $215\pm 15$ &  $+06\pm 11$ & $1.5\pm 0.4$ & $3.2\pm 0.8$ \\
14 & $S_l\geq 10$ & $|{\delta}|\leq20$ & 128952 &  $207\pm 14$ &  $+03\pm 13$ & $1.6\pm 0.4$ & $3.4\pm 0.9$ \\
15 & $S_l\geq 10$ & $|{\delta}|\leq15$ & 98022 &  $203\pm 14$ &  $-01\pm 13$ & $1.8\pm 0.5$ & $3.9\pm 1.0$ \\
\hline
\end{tabular}
}
\end{center}
\end{table*}
\begin{figure*}
\includegraphics[width=\linewidth]{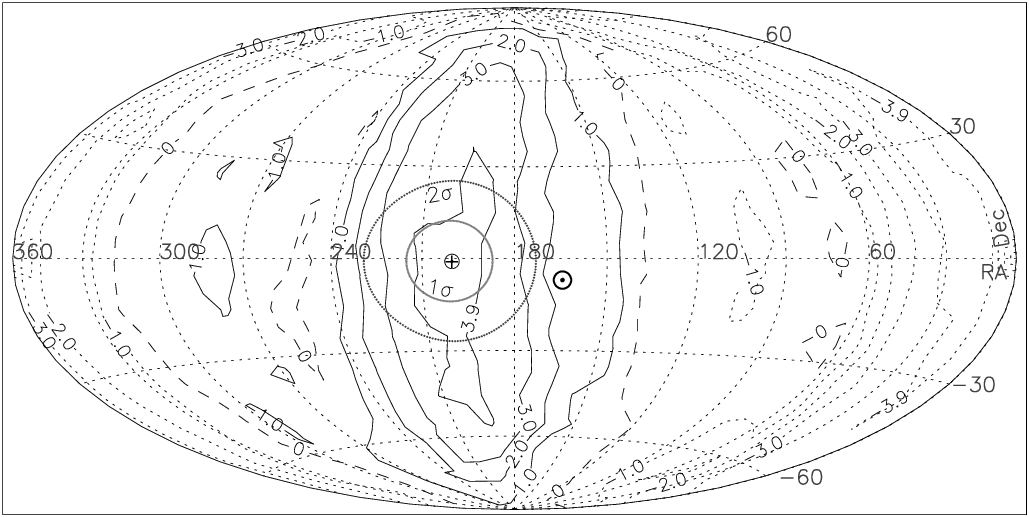}
\caption{A contour map of the dipole amplitudes, estimated for various directions in the sky, for the VLASS data restricted to $|\delta| \leq 15^\circ$ and $S_{\rm l}\geq 10$ mJy. The~horizontal and vertical axes denote RA and Dec in degrees. The~true pole direction is expected to be closer to the higher contours values, shown by continuous lines, while the true antipole should lie closer to the lower contour values, shown by dotted lines. The~dashed curve represents the zero amplitude of the dipole. 
The symbol $\oplus$ indicates the derived pole position for the VLASS sample,  along with   $1\sigma$ and $2\sigma$ error ellipses around it. The symbol $\odot$ indicates the CMB pole position. 
\label{F6}
} 
\end{figure*}
\begin{table*}
\begin{center}
\caption{\label{T5}Velocity vector from number counts for the NVSS dataset with $|b|>10^\circ$.}
\hskip4pc\vbox{\columnwidth=33pc
\begin{tabular}{ccccccccccccccc}
\hline 
(1)&(2)&(3)&(4)&(5)&(6)&(7)&(8)\\
Serial&Flux-density & Declination  & $N$ 
&\multicolumn{4}{c}{Peculiar velocity vector estimate} \\
\cline{5-8}
No.&$S$ &Coverage & &  RA & Dec & ${\cal D}$  &   $p$\\
&(mJy)&($^{\circ}$)&& ($^{\circ}$)&  ($^{\circ}$) &  ($10^{-2}$) & ($370$ km s$^{-1}$)&  \\
\hline
1 & $S\geq 50$ & $|{\delta}|\leq40$ & 91652 &  $170\pm 13$ &  $-17\pm 14$ & $2.1\pm 0.6$ & $4.2\pm1.1$ \\
2 & $S\geq 50$ & $|{\delta}|\leq30$ & 71776 &  $180\pm 14$ &  $-16\pm 15$ & $1.9\pm 0.6$ & $3.7\pm1.2$ \\
3 & $S\geq 50$ & $|{\delta}|\leq25$ & 60698 &  $184\pm 15$ &  $-14\pm 15$ & $1.9\pm 0.6$ & $3.7\pm1.3$ \\
4 & $S\geq 50$ & $|{\delta}|\leq20$ & 49394 &  $170\pm 16$ &  $-02\pm 16$ & $1.4\pm 0.7$ & $2.8\pm1.4$ \\
5 & $S\geq 50$ & $|{\delta}|\leq15$ & 37491 &  $153\pm 18$ &  $-02\pm 16$ & $1.7\pm 0.8$ & $3.3\pm1.6$ \\
\\6 & $S\geq 30$ & $|{\delta}|\leq40$ & 155110 &  $156\pm 11$ &  $-02\pm 10$ & $2.1\pm 0.4$ & $4.7\pm1.0$ \\
7 & $S\geq 30$ & $|{\delta}|\leq30$ & 121614 &  $156\pm 12$ &  $-05\pm 11$ & $1.7\pm 0.5$ & $3.8\pm1.1$ \\
8 & $S\geq 30$ & $|{\delta}|\leq25$ & 102903 &  $155\pm 13$ &  $-05\pm 12$ & $1.7\pm 0.5$ & $3.8\pm1.1$ \\
9 & $S\geq 30$ & $|{\delta}|\leq20$ & 83640 &  $137\pm 14$ &  $+01\pm 13$ & $1.8\pm 0.6$ & $4.0\pm1.3$ \\
10 & $S\geq 30$  & $|{\delta}|\leq15$ & 63536 &  $125\pm 17$ &  $+04\pm 14$ & $1.4\pm 0.7$ & $3.1\pm 1.5$ \\
\\11 & $S_l\geq 18$ & $|{\delta}|\leq40$ & 252842 &  $157\pm 10$ &  $+06\pm 10$ & $1.7\pm 0.3$ & $4.0\pm0.8$ \\
12 & $S_l\geq 18$ & $|{\delta}|\leq30$ & 198230 &  $163\pm 12$ &  $+05\pm 10$ & $1.3\pm 0.4$ & $3.1\pm0.8$ \\
13 & $S_l\geq 18$ & $|{\delta}|\leq25$ & 167701 &  $159\pm 13$ &  $+03\pm 11$ & $1.4\pm 0.4$ & $3.3\pm0.9$ \\
14 & $S_l\geq 18$ & $|{\delta}|\leq20$ & 136246 &  $149\pm 14$ &  $+05\pm 13$ & $1.6\pm 0.4$ & $3.7\pm1.0$ \\
15 & $S_l\geq 18$ & $|{\delta}|\leq15$ & 103783 &  $134\pm 16$ &  $+03\pm 15$ & $1.6\pm 0.6$ & $3.6\pm1.2$ \\
\hline
\end{tabular}
}
\end{center}
\end{table*}
\begin{table*}
\begin{center}
\caption{\label{T6}Velocity vector from number counts for the TGSS dataset with $|b|>10^\circ$.}
\hskip4pc\vbox{\columnwidth=33pc
\begin{tabular}{ccccccccccccccc}
\hline 
(1)&(2)&(3)&(4)&(5)&(6)&(7)&(8)\\
Serial&Flux-density & Declination  & $N$ 
&\multicolumn{4}{c}{Peculiar velocity vector estimate} \\
\cline{5-8}
No.&$S$ &Coverage & &  RA & Dec & ${\cal D}$  &   $p$\\
&(mJy)&($^{\circ}$)&& ($^{\circ}$)&  ($^{\circ}$) &  ($10^{-2}$) & ($370$ km s$^{-1}$)&  \\
\hline
1 & $S\geq 250$ & $|{\delta}|\leq53$ & 99736 &  $168\pm 13$ &  $-08\pm 13$ & $4.8\pm 0.5$ & $10.0\pm1.1$ \\
2 & $S\geq 250$ & $|{\delta}|\leq30$ & 64378 &  $161\pm 14$ &  $-07\pm 13$ & $5.0\pm 0.6$ & $10.5\pm1.3$ \\
3 & $S\geq 250$ & $|{\delta}|\leq25$ & 54684 &  $159\pm 15$ &  $+02\pm 14$ & $4.9\pm 0.7$ & $10.3\pm1.4$ \\
4 & $S\geq 250$ & $|{\delta}|\leq20$ & 44184 &  $155\pm 17$ &  $+05\pm 15$ & $4.9\pm 0.8$ & $10.3\pm1.6$ \\
5 & $S\geq 250$ & $|{\delta}|\leq15$ & 33069 &  $157\pm 18$ &  $+01\pm 16$ & $6.0\pm 0.9$ & $12.7\pm1.8$ \\\\
\\6 & $S\geq 150$ & $|{\delta}|\leq53$ & 161653 &  $164\pm 10$ &  $-01\pm 09$ & $4.4\pm 0.4$ & $10.6\pm1.0$ \\
7 & $S\geq 150$ & $|{\delta}|\leq30$ & 104787 &  $152\pm 13$ &  $-07\pm 12$ & $4.7\pm 0.5$ & $11.4\pm1.3$ \\
8 & $S\geq 150$ & $|{\delta}|\leq25$ & 89003 &  $151\pm 14$ &  $+02\pm 13$ & $4.8\pm 0.5$ & $11.6\pm1.3$ \\
9 & $S\geq 150$ & $|{\delta}|\leq20$ & 71793 &  $148\pm 15$ &  $+04\pm 14$ & $4.6\pm 0.6$ & $11.2\pm1.5$ \\
10 & $S\geq 150$ & $|{\delta}|\leq15$ & 53965 &  $152\pm 65$ &  $+01\pm 15$ & $5.6\pm 0.7$ & $13.7\pm1.7$ \\\\
\\11 & $S_l\geq 70$ & $|{\delta}|\leq53$ & 297088 & $151\pm 09$ &  $+14\pm 10$ & $4.2\pm 0.3$ & $10.2\pm0.8$ \\
12 & $S_l\geq 70$ & $|{\delta}|\leq30$ & 193280 &  $138\pm 15$ &  $-01\pm 12$ & $3.6\pm 0.4$ & $08.8\pm0.9$\\
13 & $S_l\geq 70$ & $|{\delta}|\leq25$ & 164232 &  $135\pm 16$ &  $+06\pm 12$ & $3.7\pm 0.4$ & $09.1\pm1.0$\\
14 & $S_l\geq 70$ & $|{\delta}|\leq20$ & 132470 &  $136\pm 16$ &  $+07\pm 13$ & $3.7\pm 0.5$ & $09.1\pm1.1$\\
15 & $S_l\geq 70$ & $|{\delta}|\leq15$ & 99968 &  $137\pm 17$ &  $+01\pm 14$ & $4.2\pm 0.5$ & $10.2\pm1.3$ \\
\hline
\end{tabular}
}
\end{center}
\end{table*}

A detailed examination of the RACS data northward of $\delta=25^\circ$, in steps of 0.05 degree in declination, reveals that at almost all flux-density levels, there is a sharp drop in the number density of sources at $\delta \geq 29\!\stackrel{^\circ}.\!\!1$, in fact there are no sources in the catalogue beyond $\delta> 29\!\stackrel{^\circ}.\!\!15$ at any flux-density level.
It does seem that large, apparently discordant, declination values for pole ($+42^\circ$ for the VLASS and $-45^\circ$ for the  RACS) seen in Singal (2023) were quite likely due to some missing sources near the declination limits in the survey data, which we hope to correct for here by employing declination restrictions on the data employed for the dipole estimates.

The results for the dipoles for the four surveys, at different flux-density levels for each sample, are presented in Tables~\ref{T3} to ~\ref{T6}, where various columns are arranged as follows.  Column (1) gives the serial number of each entry, column (2) lists the flux-density limit of the sub-sample, column (3) gives the declination-coverage range, column (4) lists the number of sources in the sub-sample, columns (5) and (6) list the direction of the dipole in terms of right ascension and declination, derived from the vector dipole method applied to that sub-sample, column (7) gives $\cal D$, the weighted mean of the dipole magnitude values  computed from Eqs.~(\ref{eq:5}) and ~(\ref{eq:6}), and column (8) lists  $p$, the peculiar speed (in units of the CMB value 370 km s$^{-1}$) estimated from $\cal D$ using Eq.~(\ref{eq:7}) for the corresponding sub-sample.
In Tables~\ref{T3} to ~\ref{T6} we have determined the dipole not only at the flux-density level $S_l$, determined from the horizontal line in Fig..~\ref{F1}, but also at some higher flux-density levels to check for any anomalies in the dipole determination. 

The results for the dipole, determined from the anisotropy in number counts in the RACS sample  at three different flux-density levels, are presented in Table~\ref{T3},  
where we see that the direction as well as amplitude of the dipole is almost independent of the chosen flux-density levels. Corresponding to the declination limit, $\delta \leq 30^\circ$, mentioned for the RACS survey (Hale et al. 2021), using  $|\delta| \leq 30^\circ$, at all  flux-density levels, the derived dipole is not only of a large amplitude, $p \geq 8.4$, but is also significantly away, especially in declination, from the CMB dipole at \mbox{RA$=168^{\circ}$}, Dec$=-7^{\circ}$. However, as soon as we change the declination limit of our RACS sample from $|\delta| \leq 30^\circ$ to $|\delta| \leq 29^\circ$, the dipole derived seems suddenly to change position in declination by about $30^\circ$ northward, towards the CMB pole. Even the dipole amplitude $p$ reduces almost by a factor of two. As we restrict the declination limit of the sample to $|\delta| \leq 25^\circ$ and continue to restrict still further in steps of $5^\circ$, the shift in dipole position in the sky or any further changes in its amplitude are much smaller.

The dipole amplitude distribution across the sky for the RACS data is shown in a contour map (Fig.~\ref{F5}) for  the sub-sample with $|\delta| \leq 15^\circ$ and $S_{\rm l}\geq 25$ mJy. 
The derived direction for the peculiar motion is indicated in Fig.~\ref{F5} by the symbol $\oplus$, along with $1\sigma$ and $2\sigma$ error ellipses. The RACS dipole direction in sky is within $2\sigma$ of the CMB pole position, indicated by the symbol $\odot$ on the map. With a  declination coverage up to $+41^\circ$ (Hale et al. 2021)  or further up to $+51^\circ$ (McConnell et al. 2020), one should have much smaller errors in both direction and amplitude of the dipole,  

\begin{figure*}
\includegraphics[width=12cm]{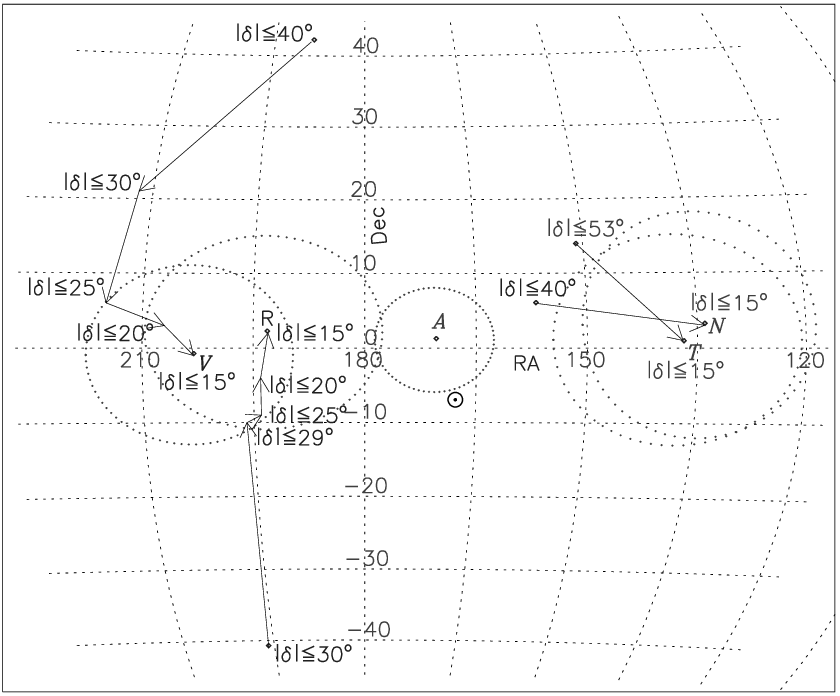}
\caption{Shifts in the positions of~the poles for the radio dipoles along with their final $1\sigma$ error ellipses, for various declination limits, from Tables~\ref{T3}-\ref{T6}. To minimize cluttering for the NVSS and TGSS dipoles, where the total shift in positions are  relatively much smaller, we have plotted only the initial and final declination limit values from Tables~\ref{T5}-\ref{T6}. Also shown is the weighted average position (indicated by $A$) along with its error ellipse, for all the four radio poles. The average dipole position $A$ seems to lie very close to the CMB pole position, at RA$=168^{\circ}$, Dec$=-7^{\circ}$, indicated by $\odot$, which has negligible errors. \label{F7}}
\end{figure*}

The results for the dipole, determined from the anisotropy in number counts in the VLASS sample  at three different flux-density levels, are presented in Table~\ref{T4}, where we see that the direction of the dipole is not always independent of the chosen flux-density levels, especially for the larger declination limits ($|\delta| \leq 40^\circ$ or $|\delta| \leq 30^\circ$) for the sample. Only as we choose  smaller declination limits ($|\delta| \leq 25^\circ$) for the sample, the dipole position not only matches at different flux-density levels but it also moves towards the CMB dipole position  (\mbox{RA$=168^{\circ}$}, Dec$=-7^{\circ}$). The dipole amplitude $p$, however, does not vary much with the declination limit of the sample or the flux-density limit of the chosen sample.

Figure~\ref{F6} shows a contour map of the dipole amplitude distribution across the sky for the VLASS data for the $|\delta| \leq 15^\circ$ and $S_{\rm l}\geq 10$ mJy sub-sample. 
The derived direction for the peculiar motion is indicated in Fig.~\ref{F6} by the symbol $\oplus$, along with the $1 \sigma$ and $2 \sigma$ error ellipses. The VLASS dipole position is $\sim 2.6\: \sigma$ away from the CMB pole position, indicated by the symbol $\odot$ on the map. 
It has already been pointed out in VLASS data (Singal 2023) that even at a declination limits ($|\delta| \leq 40^\circ$) for the sample, at $\geq 30$ mJy levels the direction of the dipole might appear to be in agreement with the CMB dipole (\mbox{RA$=168^{\circ}$}, Dec$=-7^{\circ}$), but as we go to the lower flux-density levels ($\geq 10$ mJy) the direction of the determined dipole shifts significantly away from the CMB dipole, especially in declination.
For example, for the $S>5$ mJy VLASS sample (not entered in Table~\ref{T6}), the derived dipole declination is found to be $\sim 68^{\circ}$ while for the $S>3$ mJy VLASS sample the derived dipole declination turns out to be $\sim 76^{\circ}$. This indicates that the problem of the incompleteness of the VLASS sample might be more pronounced at lower flux-density levels. The catalogue used here is derived from Quick Look images of the epoch 1 data (Gordon et al. 2021), carried out at Karl G. Jansky Very Large Array (Lacy et al. 2020) and hopefully in the second and third epoch data of the VLASS survey, these disparities at low flux-density levels will disappear or at least get sufficiently reduced to yield dipoles with sky positions much more consistent at various flux-density levels and thus more reliable. 

The agreement of the dipole values, in both direction and amplitude, for the RACS and the VLASS  samples, especially in the rows 15 of Table~\ref{T3} and Table~\ref{T4} lends credence to our contention that by restricting the data somewhat away from the regions in the vicinity of the declination limit of the survey, one can mitigate the effects of such systematics to some extent in the dipole determination. 

The results for the dipole, determined from the anisotropy in number counts in the NVSS sample  at three different flux-density levels, are presented in Table~\ref{T5}, where we see that both the direction and amplitude of the dipole varies significantly neither with the chosen flux-density levels, nor with the declination limits of the sample. The dipole position within errors matches the CMB dipole position, but the dipole amplitude $p$, however, is about $\sim 4$ times larger.

Table~\ref{T6} lists dipole, derived for the TGSS sample at different flux-density levels for  different declination limits. The dipole direction within errors agrees with the CMB dipole, the amplitude $p$, however, is about $\sim 10$ times larger, $\sim 2.5$ larger than even the NVSS dipole seen in Table~\ref{T5}.
\section{Combining the dipole vector results from various radio surveys}\label{S6}

\begin{figure*}
\includegraphics[width=15cm]{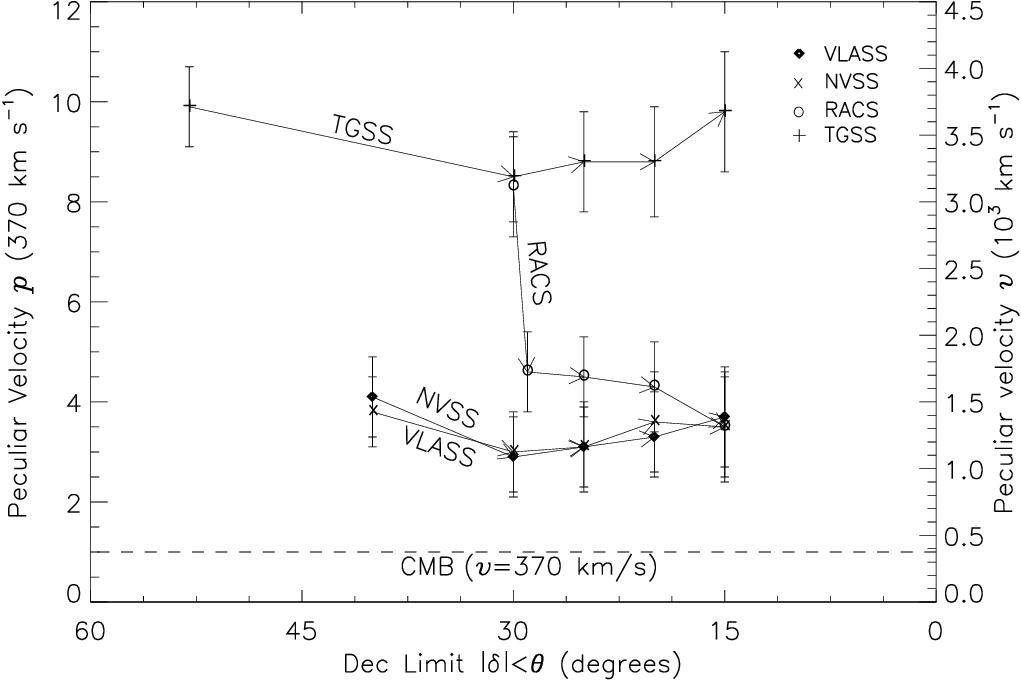}
\caption{Changes in the amplitudes of~the poles for various declination limits for the radio dipoles. There is an abrupt change in the dipole amplitude for the RACS data when the declination limit is changed from $|\delta| \leq 30^{\circ}$ to $|\delta| \leq 29^{\circ}$, which corresponds to the effect of drop in number density at the survey declination limit of $\delta = +30^{\circ}$ (Section~\ref{S4}, Fig.~\ref{F3}). \label{F8}}
\end{figure*}
Figure~\ref{F7} shows the shifts in sky positions of radio dipoles along with their final $1\sigma$ error ellipses, from Tables~\ref{T3}-\ref{T6}, for various declination limits. To minimize cluttering for the NVSS and TGSS dipoles, where the total shifts in positions are  relatively much smaller, we have plotted only the initial and final declination limit values from Tables~\ref{T5}-\ref{T6}. The CMB pole, at RA$=168^{\circ}$, Dec$=-7^{\circ}$, indicated by $\odot$, has negligible errors. 
From Fig.~\ref{F7} we see that there is quite some spread in right ascension in the radio dipole positions about the CMB dipole, whether it is caused by some unknown systematics in right ascension is not very clear. For instance,  the entries 11 to 15 in Table 5 show that as the declination limit is reduced from $|\delta|<40$ to $|\delta|<15$, RA of the NVSS dipole does not reduce systematically. In fact, RA first shifts to slightly higher values and then there is a sudden drop. Nonetheless, a weighted average of all four radio dipole positions gives for the mean dipole position RA$=170^\circ\pm 8^\circ$, Dec$=1^\circ\pm 7^\circ$, shown in Fig.~\ref{F7} as $A$, which is within $1.2\sigma$ of the CMB dipole position (RA$=168^\circ\pm 8^\circ$, Dec$=-7^\circ$) in sky. It should be noted that no a priori information about the CMB dipole position was used in the radio dipole determinations. So their weighted mean coming close to the CMB dipole position may be of a genuine significance.

Figure~\ref{F8} shows changes in the amplitudes of dipoles for various declination limits for the four radio surveys. There is an abrupt change in the dipole amplitude for the RACS data when the declination limit is changed from $|\delta| \leq 30^{\circ}$ to $|\delta| \leq 29^{\circ}$, indicating the effect of drop in number density at the survey declination limit of $\delta = +30^{\circ}$ (Section~\ref{S4}, Fig.~\ref{F3}). In three cases, VLASS, RACS and NVSS, the dipole amplitude converges to a single value, however the TGSS dipole remains $\sim 2.5$ times larger than the other dipoles.
A weighted mean of all four radio dipole amplitudes gives an average value $p=5.0\pm 0.6$, corresponding to a peculiar velocity $v=1830\pm 210$ km s$^{-1}$. However, because  the TGSS dipole amplitude differs from the average much more than five standard deviations, we may as well discard it while determining the mean. In that case we get a weighted mean of all other three  radio dipole amplitudes as $p=3.7\pm 0.6$, implying a peculiar velocity $v=1380\pm 230$ km s$^{-1}$, a value that departs from the CMB value at $\sim 4.4\sigma$ level. 

Our mean radio dipole position estimated
from the four radio surveys is in excellent agreement with that derived from a combined RACS and NVSS dataset by Wagenveld et al. (2023), who applied the Bayesian estimators, though our derived mean amplitude for the dipole ($p=3.7\pm 0.6$) is somewhat higher than the value ($\sim 3$) derived by them. 
\section{Anomaly of the much larger amplitude for the TGSS dipole}\label{S5}
\begin{table*}
\begin{center}
\caption{\label{T7}Dipole estimates for the TGSS sample for various  flux-density (S) ranges and sky-coverage limits in right ascension (RA), declination (Dec), galactic latitude (b) and supergalactic latitude (SGB).} 
\hskip4pc\vbox{\columnwidth=33pc
\begin{tabular}{ccccccccccccccc}
\hline 
(1)&(2)&(3)&(4)&(5)&(6)&(7)&(8)&(9)&(10)\\
Serial&Flux-density&\multicolumn{2}{c}{Sky-coverage range} & |b|  & |SGB|  & N&\multicolumn{3}{c}{Peculiar velocity vector estimate} \\
\cline{3-4}
\cline{8-10}
No.&(S) &RA & Dec & limit  & limit  & & RA & Dec &$p$ \\
&(mJy)&($^{\circ}$)&  ($^{\circ}$)&($^{\circ}$)&($^{\circ}$)&&($^{\circ}$)&  ($^{\circ}$)&($370$ km s$^{-1}$)\\
\hline
1&$S\geq 250$  &0-360 & $|{\delta}|\leq53$ &$>10$ & $>0$ & 99736 &  $168\pm 13$ &  $-08\pm 13$ & $10.0\pm1.1$ \\
2&$250 >S \geq 70$&0-360 & $|{\delta}|\leq53$ & $>10$ & $>0$ & 197352 & $138\pm 11$ &  $+28\pm 13$ & $11.8\pm1.1$ \\
3&$250 >S \geq 100$&0-360 &  $|{\delta}|\leq53$ & $>10$ & $>0$ & 128037 & $154\pm 12$ &  $+14\pm 14$ & $09.5\pm1.2$ \\\\
4&$S_l\geq 70$&$030-180,210-360$ & ${|\delta}|\leq53$ &$>10$ & $>0$ & 237503 &  $145\pm 12$ &  $+11\pm 12$ & $10.4\pm0.9$ \\
5&$S_l\geq 70$&$060-180,240-360$ & ${|\delta}|\leq53$ &$>10$ & $>0$ & 181752 &  $144\pm 13$ &  $+04\pm 13$ & $10.5\pm0.9$ \\
6&$S_l\geq 70$&$090-180,270-360$ & ${|\delta}|\leq53$ &$>10$ & $>0$ & 136266 &  $151\pm 14$ &  $+01\pm 14$ & $10.8\pm0.9$ \\\\
7&$S_l\geq 70$&0-360 & $10<|{\delta}|\leq53$ & $>10$ & $>0$ & 230480 & $153\pm 11$ &  $+21\pm 10$ & $09.7\pm0.9$ \\
8&$S_l\geq 70$&0-360 &  $15<|{\delta}|\leq53$ & $>10$ & $>0$ & 197120& $162\pm 12$ &  $+23\pm 11$ & $08.8\pm1.0$ \\
9&$S_l\geq 70$&0-360 & $20<{|\delta}|\leq53$ &$>10$ & $>0$ & 164618 &  $166\pm 13$ &  $+21\pm 12$ & $09.1\pm1.1$ \\\\
10&$S_l\geq 70$&0-360 &  $|{\delta}|\leq53$ &$>10$ & $>0$ &297088 &  $151\pm 09$ &  $+14\pm 10$ &  $10.2\pm0.8$\\
11&$S_l\geq 70$&0-360 & $|{\delta}|\leq53$ &  $>15$ & $>0$ &266374 & $156\pm 10$ &  $+14\pm 10$ &  $09.9\pm0.8$ \\
12&$S_l\geq 70$&0-360 &  $|{\delta}|\leq53$ & $>20$ & $>0$ &237178& $161\pm 11$ &  $+15\pm 11$ &  $09.1\pm0.8$ \\\\
13&$S_l\geq 70$&0-360 &  $|{\delta}|\leq53$ & $>10$ & $>05$ &272828& $150\pm 10$ &  $+10\pm 10$ &  $10.2\pm0.8$ \\
14&$S_l\geq 70$&0-360 &  $|{\delta}|\leq53$ & $>10$ & $>10$ &248804& $146\pm 11$ &  $+05\pm 11$ &  $08.9\pm0.9$ \\
15&$S_l\geq 7$0&0-360 &  $|{\delta}|\leq53$ & $>10$ & $>15$ &224966& $146\pm 12$ &  $-00\pm 11$ &  $08.8\pm0.9$ \\\\
\hline
\end{tabular}
}
\end{center}
\end{table*}
An issue that requires further investigation is the anomalously larger amplitude for the TGSS dipole (Table~\ref{T6}, Fig.~\ref{F8}), which is $\sim2.5$ times the value from the other three radio dipoles. It can happen that for the same peculiar velocity, much larger values of the spectral index and/or of the index of the integral source counts, could yield a larger dipole amplitude ${\cal D}$. For instance, instead of  $\alpha \sim 1$ and  $x \sim 1$ which give  ${\cal D} \sim 4 v/c$ (Eq.~(\ref{eq:7})), $\alpha \sim 2$ and  $x \sim 2$ would yield a value of  ${\cal D}$ which is twice as large, for the same value of the peculiar speed $v$. However for the TGSS data, $x$ lies between $0.8$ and $1.03$ (Fig.~\ref{F1}), while the spectral index varies between $\alpha=0.76$ and $\alpha=0.8$ (Tiwari 2019; Singal 2019a). So a factor of 2.5 times higher value of  ${\cal D}$ for the TGSS dipole could not be explained that way. 

We have determined dipole with different flux-density limits in rows 1 to 3 of Table~\ref{T7}, the dipole amplitude remains high in each case. In fact, row 1 and 2, with $S\geq 250$ mJy and $250 >S \geq 70$ mJy respectively or row 1 and 3, with $S\geq 250$ mJy and $250 >S \geq 100$ mJy respectively have no overlaps in flux density and thus no common sources. Thus from independent set of sources without any overlap, the determined dipoles are of similar high amplitudes.

Another possibility for the high amplitude could be if there were large scale systematic number-density variations in right ascension, which might not be visible in Fig.~\ref{F4}. 
To see if any such density variations in right ascension play a role, we have used in row 4 to 6 of Table~\ref{T7}, data with different symmetric cuts in right ascension, but no significant effect on the dipole amplitude is noticeable.

It has been pointed out by Secrest et al. (2022) that the high value of the TGSS dipole could be a result of declination-dependent calibration problems. But as discussed in Singal (2023), it does not seem that the observed high value for the TGSS dipole owes to such calibration errors. This is further  confirmed here as choosing different declination-coverage ranges, inferred TGSS dipole amplitudes  remain almost unchanged at various flux-density levels (Table~\ref{T6}). 

Since in Table~\ref{T6} we used declination bands centred around the equator, we can also try sky slices at high declinations by dropping bands of declination around the equator.  
Table~\ref{T7} shows sky slices away from the equator after dropping bands of low declination sources with $|\delta|\leq10^\circ$ or $|\delta|\leq15^\circ$  or $|\delta|\leq20^\circ$, and employing instead the sky slices with $10^\circ<{|\delta}|\leq53^\circ$, $15^\circ<{|\delta}|\leq53^\circ$ and $20^\circ<{|\delta}|\leq53^\circ$; there seems hardly any significant change in thus computed dipole amplitude. 
The TGSS dipole amplitude remains high for all declination cutoffs (Table~\ref{T7}).  

In fact, the sky-coverage limit $20^\circ<{|\delta}|\leq53^\circ$ in the 9th row of Table~\ref{T7}, has no overlap with the sky-coverage limit, ${|\delta}|\leq 20^\circ$ in the 14th row of Table~\ref{T6}, where, the dipole amplitudes, without any common sources, turn out to be  equally high in both cases. Also, 8th row of Table~\ref{T7} and 15th row of Table~\ref{T6} have no overlap in the sky coverage and thus have no common sources, yet the dipole amplitudes have similar high values in either case.

Could it be that influence of the galactic sources is making the dipole amplitude high at the TGSS survey frequency (150 MHz)?  In fact increasing the Galactic latitude cuts up to $15^\circ$ or even  $20^\circ$ made no significant changes in the dipole amplitudes, which remained $\sim 10$. Similar is the result for different Supergalactic latitude cuts, the TGSS dipole amplitude remains high irrespective of the various cuts made. 

It is evident that the larger amplitude of the TGSS dipole is not a result of some small pockets of excessive number density towards the pole direction or some gaps or even ``voids'' of low frequency (around 150 MHz) sources toward the anti-pole direction, since mutually exclusive regions of sky with no overlaps (cf. row 8 or 9 of Table~\ref{T7} and row 14 or 15 of Table~\ref{T6}) and thus without any common sources have yielded similar large amplitudes for the dipole. Thus, dipole amplitude for various sky-coverage limits in right ascension, declination, galactic latitudes and supergalactic latitudes for the TGSS sample remains steadfast around $p\sim 10$, an order of magnitude larger than the CMB dipole, and $\sim 2.5$ times the value from the other three radio surveys. 
\section{A peculiar motion of Solar system not conversant with various cosmic dipoles -- implications for the CP}\label{S7}

With more and more radio dipoles getting determined from anisotropies in the sky distribution of radio sources in large surveys like VLASS and RACS, it is becoming increasingly clear that a Solar peculiar motion does not fit all these observations. Now we have four independent radio surveys, VLASS, RACS, NVSS and TGSS, that have given dipoles in the same direction as the CMB dipole.
While three of the radio dipole, VLASS, RACS and NVSS, yield amplitudes consistently $\sim 3.7$ times higher than the CMB dipole, the remainder, TGSS dipole, shows an amplitude which seems higher than the other three by a factor $\sim 2.5$, which whatever possible attempts could not bring down to the amplitude level consistent with other three dipoles. Such a statistically significant difference in the the magnitude of the TGSS dipole from  other three dipoles is puzzling.

However, one thing is clear that all these dipole amplitudes are way above the CMB dipole value, 
and the evidence seems irrefutable that the peculiar velocity of the Solar system estimated from distant radio source distributions in sky is not congruent with that inferred from the CMB sky distribution, and this would not happen for a genuine peculiar velocity of the Solar system. 
Even though the directions of the dipoles are matching with that of the CMB, a significant difference in their amplitudes from the CMB raises serious doubts on the interpretation of these dipoles, including for the CMB one, in terms of a peculiar motion of the Solar system. 
In such a case, where the peculiar motion of the Solar system does not simultaneously explain  {\em various observed}  dipoles, some other possible cause may need to be investigated for the dipoles, with a common direction in the Universe inherently preferred. Here it should be noted that the dipole amplitudes found by Secrest et al. (2021,22) and Kothari et al. (2022) for the mid-infrared sources (${\cal D} \sim 1.5-1.7$) are similar to three of our radio dipoles (Tables~\ref{T3} to \ref{T5}). However, when interpreted in terms of the solar peculiar motion, because of a large value of the index of the integral source counts of  mid-infrared source population ($x\sim 1.9$) as compared to the radio source population ($x\sim 1$), translate to a peculiar velocity (Eq.~(\ref{eq:7}), Section \ref{S2}) somewhat smaller than that from the radio dipoles but still twice as large as estimated from the CMB dipole.  

According to the Cosmological Principle, the Universe is inherently isotropic, meaning similar in all directions which implies that no preferred direction would be seen by an observer that might be stationary with respect to the expanding cosmic fluid. The dipole anisotropy seen in the CMB, which otherwise showed an isotropic distribution on large scale, had a natural interpretation that this dipole is a result of our peculiar motion along with the Solar system. With no other observations that were  contradicting, this interpretation soon turned into a `folklore'. However, now we have many more dipoles, mostly inferred from anisotropies in the sky distribution of distant radio galaxies and quasars, that have repeatedly shown values of the peculiar velocity many times larger than that from the CMB, though surprisingly, in all cases the directions agreed with the CMB dipole.
A real solar peculiar velocity should be the same irrespective of the data or the technique employed to estimate it. Therefore, such discordant dipole amplitudes might mean that the `true'  explanation for these dipoles, including that of the CMB, might in fact be something else. 

For one thing, discrepancy between radio dipoles and the CMB dipole, cannot be merely due to something amiss in the radio data or techniques, otherwise these should have ended up pointing in random directions in sky. A common, unique direction in sky being picked by all dipoles, the CMB dipole included, if not due to a Solar peculiar motion, hints that there is something else `peculiar' in that particular direction which is not resulting from observer's motion in that direction. 
This suggests instead, a preferred direction in the Universe, which has all but one implication of an inherent anisotropy in the cosmos, which would be in contravention of the conventional wisdom that the Universe is isotropic on a sufficiently large scale, \`a la CP, the basis of the modern cosmology, including the standard $\Lambda$CDM model. 
\section{Conclusions}\label{S8}
Incongruency in dipoles from four large radio surveys, VLASS, RACS, NVSS AND TGSS, having the same sky coverage and with a similar source number density in sky, was shown to result from declination-dependent systematics in the radio data in the vicinity of the declination limit of some of the surveys. When the dipoles were estimated employing a smaller declination coverage of the sky, away from the declination limits of the surveys, the systematics got reduced and the inferred dipoles thereby, became increasingly congruent. A weighted mean of thus estimated four radio dipole positions gave an average dipole position lying within $1.2\sigma$ of the CMB dipole position. Amplitudes of the dipoles from three surveys, viz. VLASS, RACS and NVSS, gave a value $\sim 3.7$ times the CMB dipole. Only the TGSS dipole gave a much larger amplitude, $\sim 10$ times the  CMB dipole. None of the radio source dipole gave an amplitude similar to that of the CMB, which not only puts in doubt a Solar peculiar motion being the ultimate cause responsible for the genesis of these dipoles, especially for the CMB dipole, which according to the conventional wisdom, is a herald archetype for the Solar peculiar velocity. However, all these dipoles being along the same direction, if not due to observer's peculiar motion, indicate a preferred, unique direction, inherently present in the Universe. A scenario like this is not conversant with the CP, the basis of the modern cosmology including the standard $\Lambda$CDM cosmological model.
\section*{Data Availability}
The VLASS data used in this article are available in VizieR Astronomical Server in the public domain at http://vizier.u-strasbg.fr/viz-bin/VizieR. The dataset is downloadable by selecting catalog: J/ApJS/255/30/comp. Another, independent version of the VLASS catalog can be found in the electronic edition of the Astrophysical Journal in FITS format at https://iopscience.iop.org/article/10.3847/1538-4357/abf73b/meta. 
The RACS catalog is available at https://doi.org/10.25919/8zyw-5w85 under Files as the file  RACS\_DR1\_Sources\_GalacticCut\_v2021\_08.xml.
\section*{Declarations}
The author declares no conflicts/competing interests and no fund, grant, or other support of any kind was received for this research.

\end{document}